\documentclass[aps,prd,showpacs,nofootinbib,floats,floatfix,preprintnumbers,groupedaddress,twocolumn]{revtex4}

\usepackage{bm}
\usepackage{latexsym}
\usepackage{dcolumn}
\usepackage{amsmath,amsfonts,amssymb}
\usepackage{graphicx,epsfig}
\usepackage{amsmath}
\usepackage{fancyhdr}
\usepackage{hyperref}
\usepackage{graphicx,epstopdf}
\usepackage{dsfont}
\hypersetup{
	colorlinks   = true, 
	urlcolor     = blue, 
	linkcolor    = blue, 
	citecolor   = red 
}

\def\no{\nonumber}
\def\a{\alpha}
\def\b{\beta}

\def\d{\delta}

\def\p{\partial}

\def\na{\nabla}
\def\0{\textbf{0}}
\def\lie{\pounds_{\xi}}
\def\G{\Gamma}

\begin{document}
\title{Noncommutative Heisenberg algebra in the neighbourhood of a generic null surface}
\author{Krishnakanta Bhattacharya\footnote {\color{blue} krishnakanta@iitg.ernet.in}}
\author{Bibhas Ranjan Majhi\footnote {\color{blue} bibhas.majhi@iitg.ernet.in}}

\affiliation{Department of Physics, Indian Institute of Technology Guwahati, Guwahati 781039, Assam, India}

\date{\today}

\begin{abstract}
We show that the diffeomorphisms, which preserve the null nature for a generic null metric very near to the null surface, provide {\it noncommutative} Heisenberg algebra. This is the generalization of the earlier work (Phys. Rev. D95, 044020 (2017)) \cite{Majhi:2017fua}, done for the Rindler horizon. The present analysis revels that the algebra is very general as it is obtained for a generic null surface and is applicable for any spacetime horizon. Finally using these results, the entropy of the null surface is derived in the form of the Cardy formula. Our analysis is completely {\it off-shell} as no equation of motion is used. We believe present discussion can illuminate the paradigm of ``gravity as an emergent phenomenon'' and could be a candidate to probe the origin of gravitational entropy. 
\end{abstract}


\maketitle

\section{Introduction}
The works of Bakenstein and Hawking \cite{Bekenstein:1973ur, Hawking:1974sw, Bardeen:1973gs} led to the conclusion that the black holes are the thermodynamic objects which has entropy, proportional to the surface area of the horizon. However, later it was shown that the entropy and temperature can be can be associated with any null surface in general relativity \cite{Parattu:2013gwa} (see also \cite{Chakraborty:2015aja, Chakraborty:2015hna}). It is an important observation in the context of ``gravity as an emergent phenomenon'' as a general null surface is not a solution of any equation of the spacetime (for an extensive review on this direction, see \cite{Padmanabhan:2009vy}).  In the absence of an acceptable theory of quantum gravity, such a phenomenon can play significant role in its microscopic nature.

One of the most important and well established fact is that the usual thermodynamics can be obtained from the microscopic structure of the system when one takes the proper limit. So it is quite obvious that there must be an underlying quantum description from which the observables are estimated. This gives rise to the statistical way of measurement. The entropy, in this case, is given by the logarithmic of the total number of accesable microstates determined by certain macroscopic parameters -- known as the Boltzmann relation.  In the case of black holes, the concept of entropy and temperature comes into the picture only when one takes the quantum effect into the account. Therefore, it should be quite natural that the quantum microstates give rise to the Bakenstein-Hawking entropy, which is regarded as the benchmark to any theory which tries to quantize the gravity. The absence of a consistent quantum theory of gravity has made it even more difficult to comprehend the possible microscopic degrees of freedom which contribute to the entropy.  

Some recent works \cite{Majhi:2011ws, Majhi:2012tf, Majhi:2012nq, Majhi:2013lba, Majhi:2014lka, Majhi:2015tpa, Bhattacharya:2017pqc, Chakraborty:2016dwb} show that the Bakenstein-Hawking entropy can also be obtained when one explores the near horizon symmetry by imposing some suitable fall-off conditions. Due to the specific choice of horizon preserving symmetry, some degrees of freedom among all gauge ones, raises to the true degrees of freedom which attribute the entropy. The generators of the conserved charges satisfy the Virasoro algebra \cite{francesco} and the entropy of the surface can be obtained from the central charge using the Cardy formula \cite{Cardy:1986ie}. In this context it must be mentioned that this idea actually originated from the work of Brown and Henneaux \cite{Brown:1986nw}, which was further developed by Carlip \cite{Carlip:1998wz, Carlip:1999cy}. Therefore, in this method, one can connect the entropy with the configuration and the associated symmetry of the surface. However, the near horizon structure and the symmetry algebra should be studied in a more rigorous manner in order to get the idea about the traits of the surface at the quantum level, realization of which might help the physicists to construct a consistent quantum theory of gravity.

Inspired by the mentioned motivation, we shall present this work for a general null-surface. This is the generalization of the earlier work \cite{Majhi:2017fua} by one of the above authors where it has been shown that the charges corresponding to the Killing horizon structure preserving conditions for the Rindler metric exhibits {\it noncommutative} type algebra. Since we know, in general, the null surface is not a solution of any equation of motion of the spacetime, it would be interesting to investigate such possibility for a general null-surface. If the same result is obtained, it would imply the results are very general and may have some deeper significance to play a crucial role in revelling the quantum nature of gravitational theories. Moreover, since null surface is a local concept, the present analysis will enlighten the ``emergent nature of gravity''. In one way the idea is the following. A local horizon (Rindler horizon) can attribute temperature and entropy. In addition, the first law of thermodynamics leads to the Einstein's equation of motion for such a horizon \cite{Jacobson:1995ab} implying that the gravity is emerged from the more fundamental theory like thermodynamics. Therefore one would be interested to investigate the null surface in this context to see the generality of these concepts.

Here we shall consider the similar boundary conditions as taken in some earlier works \cite{Majhi:2012tf, Majhi:2017fua, Chakraborty:2016dwb} and compute two different Noether potentials for the two components of the diffeomorphism vectors and obtain the conserved charges along those directions. The choice of the boundary conditions on metric coefficients are based on the fact that the structure of the null boundary does not change after perturbation near the null surface. More precisely, the only condition we impose is the location of null surface must not change. Moreover, this is imposed only on the time-radial sector of the metric -- no condition on the angular or the transverse part. In this sense, it is much weaker condition compared to the other choices, existing in literature \cite{Majhi:2013lba, Majhi:2014lka, Majhi:2015tpa, Bhattacharya:2017pqc} (see \cite{Majhi:2012st} for a complete list of references), which are imposed on the full metric. We show that the algebra between these charges, in a different basis, exhibits non-commutative algebra. Moreover, as we shall show, following the Sugawara construction one can also obtain the entropy from the Cardy formula.

To highlight the importance of our result and especially of the obtained noncommutation algebra, we want to elucidate a few facts. There are two recently developed well-known methods to quantize the gravity; one is the string theory and the other one is the loop quantum gravity. In both the approaches, the non-commutativity plays the important role in the theory. In this work, we shall show that the algebra between the conserved charges, defined on some particular basis, is non-commutative. These charges are generated due to the null horizon preserving symmetry.  Moreover, these sets of charges will also show the connection with the entropy of the surface. Therefore, our work implies the close relationship with the near horizon configuration and reveals the quantum nature in terms of the noncommutative Heisenberg algebra on the surface. Moreover, our analysis gets more justified as this kind of algebra has also been obtained in string theory \cite{Seiberg:1999vs} where the spacetime is non-commutative as well.

The whole analysis is done on a general null surface, which is not a solution of any equation of motion and, hence, all the results are {\it off-shell}. We shall exhibit the whole work in the following manner. On the succeeding section, the Gaussian null coordinates will be introduced, on which the whole analysis will be performed. Thereafter, the boundary conditions on the metric will be imposed, from which the diffeomorphism vector will be identified which preserves the near horizon structure of the of a null surface. Later on, the conserved charges will be evaluated along the different components of the diffeomorphism vector. On the same occasion, we shall also define the brackets among the charges. The subsequent section will contain the paramount of our analysis; where, firstly, the Fourier modes of the charges and the brackets will be calculated. Afterwards, choosing the different basis of the charges, the noncommutative Heisenberg algebra will be presented. We shall also define new generators of charges following the Sugawara construction and it will be shown that one of the generators follows the Virasoro algebra and gives entropy from the Cardy formula. Finally, we shall present the summary and the outlook of our analysis.

{\it Notations}: The Latin letters $a,b,$ etc. stand for all spacetime indices while $A,B$, etc. take transverse (or angular) coordinates only. Moreover, we also use $\0$ which imply a zero vector in transverse directions.

\section{Near horizon symmetry, charges, brackets and the algebra}
Due to the seminal work of Bondi, Mentzer and Sachs (BMS) \cite{Bondi:1962px, Sachs:1962wk, Sachs:1962zza} and also of Brown and Henneaux \cite{Brown:1986nw}, it is known that one can define non-local conserved charges in any theory in which local gauge symmetry is present. Here, in the context of genaral relativity, we have dealt with diffeomorphism as the local gauge symmetry and have defined the charges and brackets on a null surface. Instead of getting the Heisenberg algebra (which was the case for many earlier works; such as \cite{Afshar:2016wfy, Afshar:2016uax, Grumiller:2016kcp, Sheikh-Jabbari:2016npa, Grumiller:2016pqb, Cai:2016idg, Afshar:2016kjj,Afshar:2017okz, Setare:2016jba, Setare:2016vhy}), here we show that our set-up defines non-commutative Heisenberg algebra near the null horizon. In the following, we have briefly described about the Gaussian null coordinates (GNC) in which we have made our whole analysis. Thereafter, we have defined the charges by fixing the boundary value of the metric perturbation caused by the diffeomorphism. Subsequently, the near horizon algebra of the charges will be studied.

\subsection{Null metric in Gaussian null coordinates (GNC)}
The intent of this work is to survey the near-horizon behaviour of an arbitrary null surface. In contemplation of that goal, we shall briefly discuss about some precursory constructions in this section. To describe the neighborhood of a null-hypersurface, there exists a preferable choice of a set of adapted coordinate system called the {\it Gaussian Null Coordinates (GNC)}. For more details on how the metrics is constructed can be found in \cite{Hollands:2006rj, MORALES}.
The metric is given as
\begin{equation}
ds^2=-2r\a du^2+2drdu-2r\b_Adx^Adu+\mu_{AB}dx^Adx^B~, 
\label{METRIC}
\end{equation}
where, $r=0$ corresponds to the null surface. The same has also been addressed earlier in \cite{Moncrief:1983xua}. Besides, the metric components $\a$, $\b_A$ and $\mu_{AB}$ are the smooth functions of all the coordinates $u$, $r$ and $x^A$  and $\mu_{AB}$ is invertible. On the null surface, one can attribute a set of null vectors, one we choose as $l^a=(l^u, l^r, l^{x^A})=(1, 0, \0)$, and the other one, the complimentary null vector is chosen as $k^a=(0, -1, \0)$, so that $g_{ab}l^ak^b=-1$ is satisfied. The covariant components of these two vectors are given as: $l_a=(-2r\a, 1, -r\b_A)$ and $k_a=(-1, 0, \0)$. Note, $l^a$ is a ``Killing-like'' (not exactly a Killing vector, as one can verify easily) vector having the property that it becomes a null vector on the (null) horizon. Nevertheless, as we shall argue later, $l^a$ does not define a Killing horizon. On the contrary, $k^a$ is a null vector throughout the spacetime. Furthermore, $l^a$ can be thought of as the future-outgoing null vector and $k^a$ as the future-ingoing null vector as those two are related as $l^ak_a=-1$. With this set of the null vectors, one can give a covariant definition of the elemental surface area of the null ($d-2$)-hypersurface (with $d$ is the dimension of the whole spacetime) as $d\Sigma_{ab}=-\sqrt{h} d^{d-2}x^{A}(l_ak_b-l_bk_a)$, where, $h$ is the determinant of the induced metric of the ($d-2$)-hypersurface. For the present case, one can check that the explicit form of the elemental surface area can be determined as $d\Sigma_{ur}=-\sqrt{\mu} d^{d-2}x^A$ whereas, $d\Sigma_{uA}=\mathcal{O}(r)$ and $d\Sigma_{rA}=\0$.  In the later part of our analysis, we have defined the charges and the brackets which will be computed on the close ($d-2$)-hypersurface on the neighborhood of a null surface. Therefore, the finite non-zero contribution in those calculations will come from only the hypersurface which is transverse to the $u$ and $r$ directions i.e., only from the term containing $d\Sigma_{ur}$.

As we have mentioned several times in the paper, the metric (\ref{METRIC}) in general is not chosen as a solution of equation of motion of gravitational field governed by a particular gravity theory. In our case, the theory will be taken to be diffeomorphism invariant and the charge for this symmetry, which we shall consider later, is for General theory of relativity (GR) (this is for simplicity, but one can in principle consider any other diffeomorphism invariant gravity theory as well). Of course, a choice can be done for the metric parameters $\alpha$, $\beta_A$ and $\mu_{AB}$ using the equations of motion for $g_{ab}$; which is a subset of all allowed null surfaces. Here we are not making any such further restriction on them. So present analysis does not use any information of gravitational field equations of motion and, hence, the metric (\ref{METRIC}) in general may not be a solution of a particular gravity theory, {\it i.e.} we are considering the whole set of null surfaces. Therefore, we call this as an off-shell analysis.


\subsection{Set-up: null surface preserving diffeomorphisms and charges}
In this section, we shall calculate the charges for  the diffeomorphism  $x^a\rightarrow x^a+\xi^a$ which are chosen by the null surface preserving condition, which means {\it the location of the null surface is unchanged}. The diffeomorphism vector $\xi^a$ will be shown to have two components. Considering each component as individual vector, we shall have two different charges. The brackets among them, in order to study the algebra between the Fourier modes of the charges, will be calculated. 

Our aim in this paper is to study the algebra of the charges which are defined only for the diffeomorphism of the $u-r$ sector of the spacetime (\ref{METRIC}). For that we fix the transverse component of $\xi^a$ as zero, i.e., $\xi^A=0$. Now, the other two components of $\xi^a$ can be determined by the fact that the two components of the metric, $g_{rr}$ and $g_{ur}$, does not change along $\xi^a$ and, therefore, are Lie-transported.
Hence, from the two conditions $\pounds_{\xi}g_{rr}=0$ and $\pounds_{\xi}g_{ur}=0$, one can find out the non-zero components of the vector $\xi^a$ (see the appendix \ref{APPNEW} for detail derivation):
\begin{equation}
\xi^u=F(u, x^A); \ \ \ \ \ \xi^r=-r\p_uF; \ \ \ \ \ \ \xi^A=0~; \label{XIUP}
\end{equation}
and, hence, the covariant components are 
\begin{equation}
\ \ \ \ \ \xi_u=r\p_uF+\bar{\a}F;\ \ \ \ \ \xi_r= F; \ \ \ \ \ \ \ \ \xi_A=\bar{\b}_AF~; \label{XIDOWN}
\end{equation}
where $\bar{\a}=-2r\a$ and $\bar{\b}_A=-r\b_A$. Here $F$ is, for the moment, an unknown function which depends only on $u$ and transverse coordinates $x^A$. Note that the obtained diffeomorphism vector becomes null on $r=0$. Moreover, it is easy to verify that $\pounds_{\xi}g_{uu}=\mathcal{O}(r)$, which vanishes on the null surface. Therefore, $\xi^i$ is a Killing vector near the horizon for only the $u-r$ sector of the spacetime. Also, note that the chosen conditions are like gauge only for this sector of the metric (\ref{METRIC}), expressed in the particular coordinates. Now, we define two vectors $\xi^{\pm}$, which are along the components of the vector $\xi^a$ with the definition  $\xi^{+}=(0, \xi^r, \0)$ and $\xi^{-}=(\xi^u, 0 ,\0)$. Afterwards, we shall define all the charges and the brackets along these two vectors, $\xi^{+}$ and $\xi^{-}$.

   Before entering into the calculation, let us spend some time on understanding the meaning of our imposed boundary conditions. To construct a null metric in $d$ (spacetime) dimension, one needs $d(d-1)/2$ number of independent metric components. The metric (\ref{METRIC}) in Gaussian null coordinates accords to that and $r=0$ implies a generic null surface. The main feature of a null surface in GR is that it acts as a one way membrane in the spacetime and it blocks information from the other side. It is worthy to note that the blockage of the information is determined only by the $u-r$ sector, while the angular part of the metric does not play any role in it. Therefore, as long as one is concerned with the null features of the surface, one can only focus on the $u-r$ sector of the metric. Hence, the diffeomorphism vector has been formed in such a way that it preserves $g_{uu}$, $g_{ur}$ and $g_{rr}$ and, thereby, acting as the isometry in this sector which governs the characteristics of the null surface. The other components of the metric (for instance $g_{rA}$) might change along the diffeomorphism but, it does not affect the null characteristics or the location of the null surface (which is $r=0$). Note that initially, the null surface at $r=0$ was the induced metric of the angular coordinates (i.e. $\mu_{AB}$) and, finally after diffeomorphism, one is again left with the induced metric of angular coordinates as the $drdx^A$ part of the metric vanishes (because, although metric component $g_{rA}$ is non-zero under these conditions, but at $r=0$ it does not contribute, as $dr=0$, and hence again we obtain null surface). It must be emphasised that our condition is much weaker than treating the whole metric as diffeomorphism invariant near the null surface and sufficient to make the null character invariant. Here we shall show that this has an interesting feature. 
Moreover, it must be mentioned that same boundary condition was adopted earlier in various cases. For instance, finding the entropy associated to null surface in the context of Virasoro algebra \cite{Chakraborty:2016dwb}. Also similar one plays an important role in hydrodynamics of gravity \cite{Eling:2012xa} and membrane paradigm - horizon Bondi-Metzner-Sachs symmetry \cite{Eling:2016xlx}.
All these indicates that $\xi^a$ is not a Killing vector for the full metric. Rather $\xi^a$ appears as the Killing vector for the $u-r$ sector only. If we have imposed the condition that the whole spacetime is unaltered due to the arbitrary diffeomorphism ({\it i.e.} by considering $\pounds_{\xi} g_{uA}=\pounds_{\xi} g_{rA}=\pounds_{\xi} g_{AB}=0$ as well along with our conditions for the $u-r$ sector), then $\xi^a$ would have been a Killing vector. But, as has been mentioned earlier, that is a more strong condition to be imposed on the spacetime, whereas our condition is the minimum requirement to keep null structure invariant and certainly a much weaker one.

       To calculate the charges and construct the algebra, one has to calculate the components of the Noether potential due to the diffeomorphism symmetry. This is, for General theory of Relativity (GR), given by the anti-symmetric tensor:
\begin{equation}
J^{ab}=\frac{1}{16\pi}[\nabla^a\xi^b-\nabla^b\xi^a]~.
\end{equation} 
For proceeding further, let us discuss why the above diffeomorphims can be applied in this form of Noether potential. It is very much well known that the Einstein-Hilbert action is invariant under any diffeomorphism $x^a\rightarrow x^a+\xi^a$. Therefore, one can obtain a conserved Noether current $J^a$ from the Noether's theorem due to the mentioned diffeomorphism invariance of the action. As $J^a$ is conserved (i.e. $\nabla_a J^a=0$), one can write $J^a$ as $J^a=\nabla_bJ^{ab}$. For the GR case, the Noether potential is given above. The conserved Noether charge is then defined as $Q = \int d\Sigma_aJ^a$ over a bulk three-volume transverse to a (usually timelike) congruence. The integral can be farther expressed in terms of the surface integral (using the Gauss's law) as $Q=(1/2)\oint d\Sigma_{ab}J^{ab}$, where the surface encloses the three-volume and is compact. If there is a horizon in the spacetime, the compact surface in the Gauss's law consists of the horizon and the asymptotic infinity. Usually, the horizon  part is related to horizon entropy. More precisely, Wald showed that if one calculate this on the horizon for a timelike Killing vector and multiply it by $2\pi/\kappa$ ($\kappa$ is the surface gravity), this turns out to be the entropy of black hole (see \cite{Iyer:1994ys, Padmanabhan:2013xyr}). With this idea we define our charge as $Q=(1/2)\int d\Sigma_{ab}J^{ab}$ where the calculation will be done on the horizon (here it is on the null surface). The important point is that in this definition till now $\xi^a$ is completely arbitrary which is reflected from the fact of the symmetry of the action under any diffeomorphism.

   Now, to relate this charge to some physical quantity, one has to choose $\xi^a$ using particular condition. One such condition is that due to the diffeomorphism $x^a\rightarrow x^a+\xi^a$, all the metric components are invariant. Then that particular choice of $\xi^a$ is known as the Killing vectors. This is usually followed extensively and a subset of all possible allowed diffeomorphism which keeps the action invariant. But there is no hard and fast rule to choose them in this way; it is merely a particular choice of $\xi^a$. As we explained earlier, in our definition of charge, $\xi$ is completely arbitrary. Keeping this in mind, we provide another choice of them by imposing the condition that the null metric remains null.  As explained earlier, we find that for this the sufficient condition is it leaves the $u-r$ sector of the metric invariant. Precise meaning of such condition is-- it is the minimum criteria for the location of the null surface being unchanged. Such a choice has been taken earlier and it has interesting connection with horizon entropy \cite{Chakraborty:2016dwb} and fluid-gravity correspondence \cite{Eling:2012xa}. 
   In literature, there is another important choice of boundary condition exists which keeps the asymptotic form of the metric invariant. This set of boundary conditions gives rise to the superrotation and supetranslation algebra \cite{Donnay:2015abr}.
As already mentioned that the charge is defined for any $\xi^a$ and since there is no unique way to choose this, one has the freedom to find $\xi^a$ by different boundary condition. Of course, the choice should be such that it leads to any meaningful quantity. With this spirit, we have adopted another method to fix the gauge. We choose those particular $\xi^a$ vectors for which only the location of the null surface is unchanged. Moreover this keeps the null surface as null again. We found that our choice is the minimal condition on the metric coefficients which does not violet the nature of the surface at $r=0$ (it remains null under these boundary conditions). In this sense the present one is weaker condition than the earlier one. This symmetry is very much significant in the context of GR as the obtained diffeomorphism gives rise to the entropy \cite{Chakraborty:2016dwb} of the null horizon. Therefore, our method of fixing the gauge has no connection with those which results in the superrotation and supetranslation algebra; both are different in nature. As far as the motivation is concerned, here we shall show that gauge degrees of freedom, which raises to the true degrees of freedom due to the horizon preserving conditions and attribute to the entropy. Moreover, the conditions are imposed by the physical reason that the null surface remains null.

  Here, only $J^{ur}$ is needed to be computed to calculate the charge which is given as
\begin{eqnarray}
J^{ur}&=&\frac{1}{16\pi}[\p_r\xi^r-\p_u\xi^u+\bar{\b}^A\p_A\xi^u
\\
\nonumber
&&+(\p_r\bar{\a}-\bar{\b}^A\p_r\bar{\b}_A)\xi^u]~. \label{JUR}
\end{eqnarray}
Now, if we calculate the potential along $\xi^{+}$ and $\xi^{-}$ with the definition $J^{ur}_{(\pm)}=J^{ur}[\xi^{\pm}]$, the expressions will be given as
\begin{equation}
J^{ur}_{(+)}=-\frac{\p_uF}{16\pi}~, \label{JUR+}
\end{equation}
and 
\begin{equation}
J^{ur}_{(-)}=-\frac{1}{16\pi}[\p_uF+2\a F]+\mathcal{O}(r)~.
\label{JUR-}
\end{equation}
The calculation of the corresponding charges are very straightforward, which are defined as \cite{Majhi:2011ws}:
\begin{equation}
Q^{\pm}=(1/2)\int_{\mathcal{H}} d\Sigma_{ab}J^{(ab)}_{(\pm)}~, \label{QPM}
\end{equation}
where $\mathcal{H}$ stands for the fact that the integration has to be evaluated on the null surface $r=0$ for the metric (\ref{METRIC}).

 Let us now comment on one more fact which shall be followed throughout the analysis. We define charge as the surface integral of the Noether potential which has to be evaluated on the mentioned null surface. Now, the metric components $\a$, $\b$ and the determinant of $\mu_{ij}$ can be expanded in Taylor series as a function of $r$. For example, $\a(r,u,x^A)=\a_0(u, x^A)+r\a_0'(u, x^A)+....$ So, at the null surface, the leading order terms of these quantities, such as $\a_0, \b_0, \mu_0$, will contribute. But, for the sake of convenience, we shall keep them as $\a$, $\b$ and $\mu$. However, a prudent reader must understand that those are actually the leading order terms when they appear on the charges and later on the brackets of the charges which shall be calculated on the horizon. 
 
Following the definition of the charge \eqref{QPM}, one can obtain
\begin{equation}
Q^{+}=\frac{1}{16\pi}\int \sqrt{\mu} d^{d-2}x^A(\p_uF)~, 
\label{Q+}
\end{equation}
and
\begin{equation}
Q^{-}=\frac{1}{16\pi}\int \sqrt{\mu} d^{d-2}x^A[\p_uF+2\a F]~. 
\label{Q-}
\end{equation}

Now we need to obtain the bracket among these charges. In literature there is no unique way to define it. Among various definitions \cite{Silva:2002jq, Donnay:2016ejv}, we shall use here the following expression:
\begin{equation}
[Q_1,Q_2]:=\int_{\mathcal{H}} d\Sigma_{ab}[\xi^a_2J^b_1-\xi^a_1J^b_2]
\label{defbracket}
\end{equation}
which was obtained by one of the authors with T. Padmanabhan in a earlier work \cite{Majhi:2011ws}.  Subsequently, this has been used in several contexts (see, \cite{Majhi:2012tf,Majhi:2012nq, Majhi:2013lba, Majhi:2014lka, Majhi:2015tpa, Bhattacharya:2017pqc, Chakraborty:2016dwb, Majhi:2017fua} for example). The connection with other ways of defining the bracket has been discussed in \cite{Silva:2002jq, Majhi:2017fua}. It is obvious from the above that one needs to calculate the components of current $J^a$ which is related to $J^{ab}$ by $J^{a}_{(\pm)}=\na_bJ^{ab}_{(\pm)}$~. The explicit form of the currents are given as
\begin{equation}
J^{r}_{(+)}=\frac{1}{16\pi}[\p_u^2F+(\p_uF)\p_u(\ln\sqrt{\mu})]+\mathcal{O}(r)~,
\label{JR+}
\end{equation}
and, similarly,
\begin{align}
J^{r}_{(-)}=\frac{1}{16\pi}\Big[\p_u^2F+[2\a+\p_u(\ln\sqrt{\mu})]\p_uF
\no
\\
+[2\p_u\a+2\a\p_u(\ln\sqrt{\mu})]F\Big]+\mathcal{O}(r)~, \label{JR-}
\end{align}
where, we have used the fact that $J^{rA}_{(\pm)}=\mathcal{O}(r)$.  The relevant brackets, calculated by using (\ref{defbracket}), are found to be:
\begin{align}
[Q_1^{+}, Q_2^{+}]=\int d\Sigma_{ur}[\xi_2^uJ_{1(+)}^r-\xi_2^rJ_{1(+)}^u]-(1\leftrightarrow 2)
\no 
\\
=\frac{1}{16\pi}\int \sqrt{\mu}d^{d-2}x^A[(F_1\p_u^2F_2-F_2\p_u^2F_1)
\no 
\\
 +(\p_u \ln\sqrt{\mu})(F_1\p_uF_2-F_2\p_uF_1)]~; 
\label{Q1+Q2+}
\end{align}
\begin{align}
[Q_1^{-}, Q_2^{-}]=\int d\Sigma_{ur}[\xi_2^uJ_{1(-)}^r-\xi_2^rJ_{1(-)}^u]-(1\leftrightarrow 2)
\no 
\\
=\frac{1}{16\pi}\int \sqrt{\mu}d^{d-2}x^A[(F_1\p_u^2F_2-F_2\p_u^2F_1) \ \ \ \ \ \
\no 
\\
+(2\a+\p_u \ln\sqrt{\mu})(F_1\p_uF_2-F_2\p_uF_1)]~; 
\label{Q-Q-}
\end{align}
and
\begin{align}
[Q_1^{+}, Q_2^{-}]=\int d\Sigma_{ur}[\xi_2^uJ_{1(+)}^r-\xi_2^rJ_{1(+)}^u]
\no 
\\
-[\xi_1^uJ_{2(-)}^r-\xi_1^rJ_{2(-)}^u]
\no 
\\
=\frac{1}{16\pi}\int \sqrt{\mu}d^{d-2}x^A[(F_1\p_u^2F_2-F_2\p_u^2F_1) \ \ \ \
\no 
\\
+2\a F_1\p_uF_2+(2\p_u\a+2\a\p_u\ln\sqrt{\mu})F_1F_2]~. 
\label{Q+Q-}
\end{align}
Once again, let us remind that $\a$ and $\mu$ are the leading order contributions of the same quantities in the above equations \eqref{Q1+Q2+}, \eqref{Q-Q-} and \eqref{Q+Q-}. In the above three equations, we need not put the value of $J^u_{1(\pm)}$ and $J^u_{2(\pm)}$ as those are multiplied with $\xi_2^r$ and $\xi_1^r$ respectively, which are $\mathcal{O}(r)$.
Next we need to find the Fourier modes of the above charges and brackets by using the Fourier decomposition of the function $F$. In this process one needs to perform the integrations.

As a mindful reader has already noticed, the integrations of the equations (\ref{Q+}), (\ref{Q-}), (\ref{Q1+Q2+}), (\ref{Q-Q-}) and (\ref{Q+Q-}) are not exactly obtainable in the required form as the explicit form of the functions $\a$ and $\mu$ are not exactly known. But, one is needed to solve those for the sake of obtaining a compact near-horizon algebra. Some earlier works has been tried to compute similar integrations under some assumptions without giving the physical explanations of those assumptions. For example, in \cite{Chakraborty:2016dwb} it has been presumed that $\a$ and the transverse metric coefficients $\mu_{AB}$ are independent of the coordinate $u$ and a particular transverse coordinate, which does not correspond to any physical situation. However, in our work we shall show that considering the leading order contributions of $\a$ and $\mu$ as independent of \textit{only one} coordinate $u$ is enough for the purpose. It is needed to be mentioned that although it is assumed that $\mu$ is independent of $u$, does not require that all the components of $\mu_{AB}$ to be independent of $u$ -- only the determinant of the metric is needed to fulfil that criteria. Since we can set the ``restricted liberty'' on $\mu_{AB}$ that the components may be the functions of $u$, therefore, it can be shown that the horizon, we are interested in, is not a Killing horizon. A more general discussion, whether $l^a$ defines a Killing vector, is tested on the appendix \ref{APP2}. In this sense, the present condition is much more weaker than that taken in \cite{Chakraborty:2016dwb}.  
Under these weaker restrictions Eqs.  (\ref{Q1+Q2+}), (\ref{Q-Q-}) and (\ref{Q+Q-}) reduce to the following forms: 
\begin{align}
[Q_1^{+}, Q_2^{+}]=\frac{1}{16\pi}\int \sqrt{\mu}d^{d-2}x^A[(F_1\p_u^2F_2-F_2\p_u^2F_1)~; 
\label{RES1}
\end{align}
\begin{align}
[Q_1^{-}, Q_2^{-}]=\frac{1}{16\pi}\int \sqrt{\mu}d^{d-2}x^A[(F_1\p_u^2F_2-F_2\p_u^2F_1) \ \ \ \ \ \
\no 
\\
+2\a(F_1\p_uF_2-F_2\p_uF_1)]~; \ \ \ \
\label{RES2}
\end{align}
and, the last one
\begin{align}
[Q_1^{+}, Q_2^{-}]
=\frac{1}{16\pi}\int \sqrt{\mu}d^{d-2}x^A[(F_1\p_u^2F_2-F_2\p_u^2F_1) \ \ \ \ \ \ \ \
\no 
\\
+2\a F_1\p_uF_2]~. 
\label{RES3}
\end{align}

Before going into the next step, let us mention the possible meaning of the condition: $\mu$ is independent of $u$.
Consider an apparent horizon, which is a marginally trapped surface \cite{Nielsen:2008cr} and, hence, is determined by the conditions of the expansion parameters $\Theta^{(l)}=q^{ab}\nabla_al_b=0$ and $\Theta^{(k)} = q^{ab}\nabla_ak_b<0$, where $q_{ab} = g_{ab} + l_ak_b+l_bk_a$. Now for the present metric (\ref{METRIC}), $\Theta^{(l)}=\p_u\ln\sqrt{\mu}$ and $\Theta^{(k)} =-\p_r\ln\sqrt{\mu}$. This implies, near the null surface $r=0$, $\Theta^{(l)}$ vanishes, provided $\mu$ is independent of $u$. Also one can note that the other condition $\Theta^{(k)} <0$ is automatically satisfied. In this sense, the condition, $\mu$ is independent of $u$, implies that the near null surface geometry is a particular class of null surface, known as apparent horizon. This was not mentioned in the earlier work \cite{Chakraborty:2016dwb}.  However, the physical implication of assuming $\a$ as independent of $u$ is still left as an open problem.

In the following section we shall calculate the Fourier-modes of the charges and the brackets where we shall use these information to get the explicit values of the charges and the brackets.
\subsection{Algebra}
Now, to calculate the Fourier modes of the charges and the brackets, we first define the Fourier modes of $F$ as 
\begin{equation}
F_m=(1/a)\exp[im( au+p_Ax^A)]~,
\label{FF}
\end{equation} 
where, $m$ and $p_A$ include all the positive and negative integers. Also, the periodicity of the coordinate $u$ has been accounted, given by $R=2\pi/a$ with $a$ being a constant. Later it will be shown that this periodicity will help us to execute these intractable integrations. The detail method of solving the integrations of the Fourier modes of the charges and the brackets is described in the appendix \ref{APP} by taking the leading order term of $\mu$ and $\a$ as independent of $u$ on the horizon. From \eqref{Q+} and \eqref{Q-}, using the key results of \eqref{MAIN1} and \eqref{MAIN2}, we find: 
\begin{equation}
Q_m^{+}=0~;\ \ \ \ Q_m^{-}=\frac{C}{2}\d_{m,0}~; \ \ \ \ \ \ \ \ \ \ \label{QFOU}
\end{equation}
and \eqref{RES1}, \eqref{RES2}, \eqref{RES3} yield
\begin{eqnarray}
&&[Q_m^{+}, Q_n^{+}]:=0~;\,\,\,\  [Q_m^{-}, Q_n^{-}]:=-iCm\delta_{m+n,0}~;
\nonumber
\\
&&[Q_m^{+}, Q_n^{-}]:=-i\frac{C}{2} m\delta_{m+n,0}~;
\label{BRACFOU}
\end{eqnarray}
where $C=\a A/4\pi a$ with $A$ being the transverse surface area, as defined in the appendix \ref{APP}. 
The expression of the Fourier modes of the charges and the brackets look exactly similar to the same of the earlier work \cite{Majhi:2017fua} by one of us for the Rindler spacetime.  Here we show these are much more general -- valid even for a generic null surface in the Gaussian null coordinates.

 Now let us discuss the underlying significance of the above results. For that we need go in a new basis. This is exactly similar to the earlier work \cite{Majhi:2017fua}. We make our choice as follows:
\begin{align}
P_0=Q_0^{+}+Q_{0}^{-}, \ \ \ \   P_m=\mathds{A}Q^{+}_{-m}+\mathds{B}Q^{-}_{-m} ({\textrm{with}}~ m\neq 0)~,
\no 
\\
X_m=\mathds{C}Q^{+}_{m}+\mathds{D}Q^{-}_{m}~,\ \ \ \ \ \ \ \ \ \ \  \
\label{23}
\end{align}
where, the coefficients $\mathds{A}$, $\mathds{B}$, $\mathds{C}$ and $\mathds{D}$ have some freedom to take particular values. Here, we show that for some compulsive choice of these coefficients, we get the non-commutative algebra near the null horizon.
\vskip 1mm
\noindent
\textbf{Case 1:}
For the choice $\mathds{A}=\mathds{B}=\pm 1/(Cm)$ and $\mathds{C}=\mathds{D}=\mp 1/2$ one gets the anti-commutation algebra between $X_n$ and $P_n$, which is given as follows.
\begin{align}
[X_m, X_n]=-\frac{i(m-n)C}{4}\d_{m+n,0}~; \ \ \ \ \ \ \ \
\no 
\\
[P_m, P_n]=\frac{4i}{(m-n)C}\d_{m+n,0}~; \ \ \ \ [X_m, P_n]=i\d_{m,n}~. \label{CASE1}
\end{align}
\vskip 1mm
\noindent
\textbf{Case 2:} If one chooses $\mathds{A}=-1/(Cm)\pm(\sqrt{1+1/m^2})/C-(1/m\pm\sqrt{1+1/m^2}), \ \ \mathds{B}=1/m\pm\sqrt{1+1/m^2}, \ \ \mathds{C}=-(1+1/C)$ and $\mathds{D}=1$, one gets 
\begin{align}
[X_m, X_n]=\frac{i}{2}(m-n)\d_{m+n,0}=[P_m, P_n]~;
\no 
\\
[X_m, P_n]=i\d_{m,n}~. \ \ \ \ \ \ \ \ \ \ \ \ \ 
\label{CASE2}
\end{align}
\vskip 1mm
\noindent
\textbf{Case 3:} Another choice is being made as follows. $\mathds{A}=-2/m, \ \ \mathds{B}=2/m, \ \ \mathds{C}=-(1+1/C)$ and $\mathds{D}=1$. Then the brackets are 
\begin{align}
[X_m, X_n]=\frac{i}{2}(m-n)\d_{m+n,0}~; \ \ \ [P_m, P_n]=0~;
\no 
\\
[X_m, P_n]=i\d_{m,n}~. \ \ \ \ \ \ \ \ \ \ \ \ \ 
\label{CASE3}
\end{align}
\vskip 1mm
\noindent
\textbf{Case 4:} Lastly, we show that another choice is possible which is as follows. $\mathds{A}=2/(mC)-m/2, \ \ \mathds{B}=m/2, \ \ \mathds{C}=1$ and $\mathds{D}=-1$. Then
\begin{align}
[X_m, X_n]=0~; \ \ \ [P_m, P_n]=\frac{i}{2}(m-n)\d_{m+n,0}~;
\no 
\\
[X_m, P_n]=i\d_{m,n}~. \ \ \ \ \ \ \ \ \ \ \ \ \ 
\label{CASE4}
\end{align}
These anti-commutation relations near the null-horizon, shown in the above equations \eqref{CASE1}, \eqref{CASE2}, \eqref{CASE3} and \eqref{CASE4}, are the key results of our analysis. Note, the same was obtained earlier for the Rindler metric \cite{Majhi:2017fua} which is a subclass of (\ref{METRIC}) for specific constraint on these metric coefficients. As was for the prior case, the results shown above illustrates the ``restricted'' non-commutative algebra since the non-commutativity prevails only for the condition $m+n=0$ and not for any arbitrary $m$ or $n$.

Before going to the next stage, let us now discuss the necessity of the present analysis. Rindler frame is adopted by an uniformly accelerated observer on a Minkowski spacetime. Therefore, even if the metric coefficients are coordinate dependent, the spacetime is still inherently flat. Note, in that case (Rindler), only metric coefficients corresponding to ($u-r$) sector depend on coordinates while the transverse ones are constant as this sector is a flat plane. Although, equivalence principle accounts the role of accelerated frame in exploring gravity; but in some situation all features of the true curved spacetimes do not emerge from this simple model. For example, at the classical level the motion of a particle is non-trivially effected in presence of curved background and also in the case of Hawking radiation, which is a quantum phenomenon, the emitted spectrum is modified by the grey body factor.
On the other hand, in Rindler spacetime, one does not account these -- the emitted spectrum is purely Planckian in nature. These suggest that one needs to incorporate the true curvature in spacetime metric to know more about gravity. 

It has been a general belief that the local nature of gravity can be always interesting. Now, locally an observer can adopt a null surface, metric coefficients of which is spacetime dependent in a general scenario. Considering such situation, one can write a $d$ dimensional metric (\ref{METRIC}) in coordinates adapted to this null surface. In this case, only one restriction is that, on the null surface, the induced metric is only the ($d-2$) dimensional transverse surface. These indicate that the metric coefficients are not necessarily restricted to a Rindler one. Accelerated (or Rindler) frames can be regarded as a subset of particular kind. Another example can be more illuminating.    
The near horizon structure of a Kerr black hole, is not exactly a Rindler one \cite{Booth:2012xm}. It has a form similar to our present metric (\ref{METRIC}) with some more restrictions on the coefficients.  Therefore, any feature related to Rindler metric is not guaranteed by other null cases and, hence, those must be tested in order to find the generality of them. Therefore, it is necessary to investigate the validity of the earlier result of \cite{Majhi:2017fua} in the case of GNC.

 In our present analysis, we have taken the most general metric (in GNC) for a null surface (see \cite{ Hollands:2006rj, MORALES} for the construction of the metric) whose metric coefficients are functions of coordinates. Moreover, the metric coefficients  can depend on timelike coordinates as well. This is the crucial difference between the Rindler form and our present form. In that sense our situation is much more general. As we have shown that one can formulate the non-commutative Heisenberg algebra for the metric in GNC, the result will hold for any null surface-- including the horizon of a Kerr black hole or any other Rindler horizon as well. This indicates that the obtained property has a wide generality.
 
Here, we have made a particular type of linear combination of our primary charges (see Eq. (\ref{23})). The primary goal is to get an algebra between $X_m$ and $P_m$ such that they satisfy the Heisenberg algebra {\it i.e.} $[X_m, P_n]=i\delta_{m,n}$. Motivation for such aim is the following. In quantum mechanics, position and the corresponding momentum do not commute which plays a huge role to get a quantum description of any system. Since, till date, no successful quantum description of horizon exists in literature, it may be worthy to think in this direction. Therefore we demanded such algebra as a staring point.  
Now given this, we found that there are four possible choices of the coefficients $\mathds{A}$, $\mathds{B}$, $\mathds{C}$ and $\mathds{D}$. In all cases, the algebra is not the usual one.  We have obtained three types of anti-commutativity. Case 1 and case 2 represents the anti-commutativity in both $X_m$ and $P_m$. Case 3 represents the anti-commutativity in $X_m$ only, whereas the $P_m$ commutes. Lastly, case 4 represents the anti-commutativity in $P_m$ only, where $X_m$ commutes. In this respect the choices are not arbitrary.

Let us now understand why we have decomposed $\xi^a$ in terms of specific $\xi^+$ and $\xi^-$ vectors. The reason is the following. Remember, the spacetime (\ref{METRIC}) represents a generic null surface at $r=0$. Interestingly, the same surface is also represented by $u=$ constant. So on the null surface, we have two normals: one is along $r$ direction and other one is along $u$ direction. Both of them are null vectors on this surface. This is consistent with the general feature of null surface: one can always find an auxiliary null vector corresponding to the null vector which defines the location of the null surface.  Hence, in the present situation, we can choose a null-null basis to analyze the properties of the surface. Here one is along $r$ direction and other one is along $u$ direction and any tensorial quantity can be decomposed in these basis. This is the main motivation of our present decomposition of $\xi^a$ vector as null-null basis is quite natural to examine the behavior of null surfaces. We are interested to see the algebra of the charges which are defined along these two directions ({\it i.e.} along $\xi^+$ and $\xi^-$).  Moreover, as we shall address in the conclusion, individually $\xi^+$ and $\xi^-$ preserve the null surface structure like $\xi^a$. In this sense, the division of $\xi^a$ along $\xi^+$ and $\xi^-$ is unique. 

However, one can choose other basis like, $\xi^1=(b\xi^u, a\xi^r, \textbf{0})$ and $\xi^2=(d\xi^u, c\xi^r, \textbf{0})$ with $a+c=1$ and $b+d=1$ {\footnote{We thank the referee for bringing this point to our mind.}}. In that case the algebra of the charges along these direction can be derived from (\ref{BRACFOU}). The new charges in terms of the earlier ones are
\begin{eqnarray}
&&Q^1_m=aQ^+_m+bQ^-_m~;
\nonumber
\\
&&Q^2_m=cQ^+_m+dQ^-_m~.
\end{eqnarray}
So using (\ref{BRACFOU}), one finds the algebra among the above charges as
\begin{eqnarray}
&&\Big[Q^1_m,Q^1_n\Big]=\Big(-ab-b^2\Big)iCm\delta_{m+n,0}~;
\nonumber
\\
&&\Big[Q^1_m,Q^2_n\Big] = \Big(-\frac{a}{2}+ab-\frac{3b}{2}+b^2\Big)iCm\delta_{m+n,0}~;
\nonumber
\\
&&\Big[Q^2_m,Q^2_n\Big] = \Big(-2+3b+a-ab-b^2\Big)iCm\delta_{m+n,0}~.
\end{eqnarray}
Note that the above will  reduce to (\ref{BRACFOU}), for $a=1$ and $b=0$. Here we discuss for different values of $a,b$; i.e. the basis can be different from $r$ and $u$ directions. In this case again one can show that the combination like (\ref{23}) satisfies similar non-commutative Heisenberg algebra in the following manner.

We define,
\begin{eqnarray}
&&P_0=Q_0^1+Q_0^2=Q_0^++Q_0^-~,
\nonumber
\\
&&P_m=\tilde{A}Q_{-m}^1+\tilde{B}Q_{-m}^2=\mathds{A}Q_{-m}^++\mathds{B}Q_{-m}^-~,
\nonumber
\\
&& X_m=\tilde{C}Q_m^1+\tilde{D}Q_m^2=\mathds{C}Q_m^++\mathds{D}Q_m^-~.
\end{eqnarray}
In this case,
\begin{eqnarray}
&& \mathds{A}=\tilde{A}a+\tilde{B}c~, \ \ \ \ \ \mathds{B}=\tilde{A}b+\tilde{B}d~,
\nonumber
\\
&& \mathds{C}=\tilde{C}a+\tilde{D}c~, \ \ \ \ \ \mathds{D}=\tilde{C}b+\tilde{D}d~. \label{ABCDEF}
\end{eqnarray}
The same anti-commutative Heisenberg algebra between $X_m$ and $P_m$ can be obtained in this case as well for the four sets of choices of $\mathds{A}$, $\mathds{B}$, $\mathds{C}$ and $\mathds{D}$ once one demands, like earlier, $[X_m,P_n]=i\delta_{m,n}$. The coefficients $\tilde{A}$, $\tilde{B}$, $\tilde{C}$ and $\tilde{D}$ can be obtained by solving (\ref{ABCDEF}), which is given as
\begin{eqnarray}
&& \tilde{A}=\frac{\mathds{A}d-\mathds{B}c}{ad-bc}~, \ \ \ \ \ \tilde{B}=\frac{\mathds{A}b-\mathds{B}a}{bc-ad}~,
\nonumber
\\
&& \tilde{C}=\frac{\mathds{C}d-\mathds{D}c}{ad-bc}~, \ \ \ \ \ \tilde{D}=\frac{\mathds{D}a-\mathds{C}b}{ad-bc}~. \label{ABCDEFTIL}
\end{eqnarray}
Thus, similar to Case 1, if we want to obtain the same algebra between $X_m$ and $P_m$ as given in (\ref{CASE1}), the choices of $\tilde{A}$, $\tilde{B}$, $\tilde{C}$ and $\tilde{D}$ will be:
\begin{eqnarray}
&&\tilde{A}=\pm\frac{1}{mC}\frac{c-d}{bc-ad}~, \ \ \ \ \ \ \tilde{B}=\pm\frac{1}{mC}\frac{a-b}{ad-bc}~, 
\nonumber 
\\
&&\tilde{C}=\mp\frac{1}{2}\frac{c-d}{bc-ad}~, \ \ \ \ \ \ \tilde{D}=\mp\frac{1}{2}\frac{a-b}{ad-bc}~.
\end{eqnarray}
To obtain the algebra between $X_m$ and $P_m$ as given in (\ref{CASE2}) (Case 2), the choices of $\tilde{A}$, $\tilde{B}$, $\tilde{C}$ and $\tilde{D}$ will be:
\begin{eqnarray}
&&\tilde{A}=\frac{-\frac{1}{m}(\frac{d}{C}+c+d)\pm(\sqrt{1+\frac{1}{m^2}})(\frac{d}{C}-c-d)}{ad-bc}~,
\nonumber 
\\
&&\tilde{B}=\frac{-\frac{1}{m}(\frac{b}{C}+a+b)\pm(\sqrt{1+\frac{1}{m^2}})(\frac{b}{C}-a-b)}{bc-ad}~,
\nonumber
\\
&& \tilde{C}=-\frac{c+d+\frac{d}{C}}{ad-bc}~, \ \ \ \tilde{D}=\frac{a+b+\frac{b}{C}}{ad-bc}~.
\end{eqnarray}
Again, similar to case 3, we obtain the algebra between $X_m$ and $P_m$ as given in (\ref{CASE3}) for the following choices of $\tilde{A}$, $\tilde{B}$, $\tilde{C}$ and $\tilde{D}$.
\begin{eqnarray}
&&\tilde{A}=-\frac{2}{m}\frac{c+d}{ad-bc}~, \ \ \ \tilde{B}=-\frac{2}{m}\frac{a+b}{bc-ad}~,
\nonumber
\\
&& \tilde{C}=-\frac{c+d+\frac{d}{C}}{ad-bc}~, \ \ \ \tilde{D}=\frac{a+b+\frac{b}{C}}{ad-bc}~.
\end{eqnarray} 
Lastly, similar to case 4, we obtain the algebra between $X_m$ and $P_m$ as given in (\ref{CASE4}) for the following choices of $\tilde{A}$, $\tilde{B}$, $\tilde{C}$ and $\tilde{D}$.
\begin{eqnarray}
&& \tilde{A}=\frac{\frac{2d}{mC}-\frac{m}{2}(c+d)}{ad-bc}~, \ \ \ \tilde{B}=\frac{\frac{2b}{mC}-\frac{m}{2}(a+b)}{bc-ad}~,
\nonumber
\\
&& \tilde{C}=\frac{c+d}{ad-bc}~, \ \ \ \tilde{D}=-\frac{a+b}{ad-bc}~.
\end{eqnarray}
Note, for $a=1$ and $b=0$, we obtain $\tilde{A}=\mathds{A}$, $\tilde{B}=\mathds{B}$, $\tilde{C}=\mathds{C}$ and $\tilde{D}=\mathds{D}$ in each case which corresponds to our earlier result. This analysis indicates the results are indeed basis independent.

So far, to define $X_m$ and $P_m$, we have taken the linear combinations of the Fourier modes of the charges. Let us now take the combination of the charges according to the Sugawara construction to define the new generators in the following manner:
\begin{align}
J_m^{\pm}=\frac{1}{2C}\sum_{p}Q^{\pm}_{m-p}Q^{\pm}_{p}+imQ_m^{\pm}~.
\end{align}
Below, we get the interesting algebra of the newly defined generators as follows
\begin{align}
[J_m^{+}, J_n^{+}]=0~; \ \ \ \ \ \ \ \ \ \ \ \  \ \ \ \ \
\no 
\\
i[J_m^{-}, J_n^{-}]=(m-n)J^{-}_{m+n}+m^3C\d_{m+n,0}~;
\no 
\\
[J_m^{+}, J_n^{-}]=\frac{m^2}{2}Q_{m+n}^{-}-i\frac{m^3C}{2}\d_{m+n,0}~; 
\end{align}
The second relation in the above defines the Virasoro algebra, with the central charge($\tilde{C}$) is defined as $\tilde{C}=12 C$. Also, note that in this case $J_0^{-} = (1/2C) (Q_0^{-})^2$ and, hence, we define the entropy of the null surface as $S =(2\pi a/\a)Q_0^{-}=(2\pi a/\a)\sqrt{2CJ_0^{-}} $. Therefore one can write the entropy in terms of the Cardy ``like'' formula as 
\begin{align}
S=2\pi\sqrt{\frac{\tilde{C}J_0^{-}a^2}{6\a^2}}~.
\label{39}
\end{align}
Thus, the obtained results for the Rindler horizon of \cite{Majhi:2017fua} can be further extended for a more general null horizon. Moreover, since the null surface is a local concept, the above expression for entropy can be described as a {\it local} form of Cary formula.

In the original work of Cardy \cite{Cardy:1986ie}, the relation between the entropy and the Virasoro algebra via Cardy's formula was obtained due to the conformal symmetry in ($1+1$) dimensional Minkowski spacetime.
The finding of Virasoro like algebra in the case of black holes in the near horizon regime \cite{Carlip:1998wz, Carlip:1999cy} may be a deeper fact. In Cardy's work things happen at the quantum level while in gravity this is obtained at the classical computation. Therefore this similarity is not so obvious. Usually, application of the standard Cardy formula leads to the entropy which matches with that of horizon. Therefore, the validity of Cardy formula in curved spacetime although not clear but it works well.

In the present case, we adopted a reverse approach. Here the standard expression of the entropy has been taken into account. Then considering the algebra among the charges for null case, the central charge has been identified. It has been observed that combination of these information leads to the Cardy like expression (\ref{39}). This we call as a local version of Cardy formula for the following reasons. First of all, the choice of the null metric is a local concept as locally an observer can always perceive a null surface adapted to its frame. Secondly, the boundary conditions, which have been imposed on the metric coefficients, are local in nature as they are satisfied near the null surface. In addition, the validity of the local version of Cardy formula is may be due to the fact that one can always obtain a locally inertial frame even in the curved manifold. A similar expression of Cardy's formula, in the context of a null surface, has also been obtained in \cite{Chakraborty:2016dwb} as well in a different method. However, both the results (ours and of \cite{Chakraborty:2016dwb}) imply the astonishing robustness of Cardy's formula which was not predicted earlier.

Let us now comment on one important difference of our work from that of Carlip \cite{Carlip:1999cy}.  Although both the ideas are  equivalent, there is a subtle difference in the two approaches. In Carlip's approach, the algebra of the charge is obtained due to the asymptotic Killing symmetry of the Killing horizon. In that case, the diffeomorphism vector is an asymptotic Killing vector near the horizon for the {\it full} spacetime and, consequently, the different Fourier modes of the same charge lead to the Virasoro algebra:
\begin{equation}
i[Q_m, Q_n]=(m-n)Q_{m+n}+\frac{C}{12}m^3\delta_{m+n,0}~,
\end{equation}
where $C$ is known as the central charge, which is connected to the entropy by the Cardy's formula. On the other hand, our present null surface is not necessarily be a Killing horizon and the diffeomorphism vectors are chosen to asymptotic Killing for only ($u-r$) sector. Hence these are not Killing for the full spacetime. Consequently, the algebra of the corresponding charges are not central extended Virasoro algebra. Rather we find that a non-linear combination of them satisfies Virasoro algebra.
Moreover, we study the algebra of two different charges $Q^+$ and $Q^-$, which are defined along radial and $u$ directions both of which are corresponds to null vectors on the null surface.

Remember that the original Virasoro algebra is the bracket among the charges of fields corresponding to the conformal symmetry in ($1+1$) dimensional Minkowski spacetime. It comes after the quantization of the fields on this spacetime which also acquires this conformal symmetry. So such an algebra is a pure consequence of quantum field theory. On the contrary, in gravity similar type of algebra is purely classical result.
In our present case, the specific combination $X_m$ and $P_m$ of the base charges $Q^+_m$ and $Q^-_m$ leads to the interesting (anti-commutative) Heisenberg algebra. As it is well known that the commutation relation $[X, Y]=i\hbar$ originally comes from quantum mechanics. This is due to the inherent uncertainty in measurement of two observables $X$ and $Y$ simultaneously and it acts as a building block for the quantum nature of a system. So it is evident that both Virasoro and Heisenberg algebra are related to quantum nature of a system while their structure is different from each other.

\section{Conclusions and outlook}
The essence of any quantum theory is the commutation relations among the associate quantities present in the theory. Unfortunately, as mentioned earlier in numerous occasions, the quantum theory of gravity is yet to be developed in a consistent manner. Though the recently developed theoretical frameworks, namely the string theory and the loop quantum gravity, is currently trying to formulate a persuasive theory in the realm of quantum gravity, a convincing theory is not contrived yet. In this circumstances, people are trying to find out different ways which can shed some light on the quantum nature of gravity. We want to mention that some similar works \cite{Afshar:2016wfy, Afshar:2016uax, Grumiller:2016kcp, Sheikh-Jabbari:2016npa, Grumiller:2016pqb, Cai:2016idg, Afshar:2016kjj,Afshar:2017okz, Setare:2016jba, Setare:2016vhy} are going on along the same route of this work where the bracket algebra (formulated in the same spirit as of the commutation relation in quantum mechanics) between the conserved charges are being studied. Earlier, one of us has explored the similar non-commutative algebra on the neighborhood of the null surface in the Rindler background. Therefore, this work is the generalization to the earlier one which justifies the acceptability of the previous results in a much more general scenario. However, as shown in our analysis, when the spacetime is described using GNC, lots of complexities emerges in the picture.

Here, we have imposed the boundary conditions of the metric perturbation and have identified the set of diffeomorphism vectors $\xi^i$ that preserves the null horizon structure. 
 After that, we have computed the conserved charges along the components of the diffeomorphism vector and also have defined the brackets of the charges. Later on, the Fourier modes of the charges and brackets have been computed. On that course, we have mentioned the challenges that one faces to get a compact algebra and have mentioned the procedure to overcome it. Also, it has been clarified that the surface, which we are dealing with, is not a Killing surface even with our assumptions. As one can easily point out, the algebra shown by the brackets of the charges are insignificant. Later, we were keen to know, whether these algebras show any physical importance in different basis of the charges. Interestingly, the answer we have found is \textit{``Yes!''}. As we have shown, one can define four different basis of charges for which one gets noncommutative Heisenberg algebra. However, the noncommutativity exists only for some particular choices of the Fourier modes. Therefore, one can say the algebra can be categorized as a \textit{``restricted''} class of noncommutative Heisenberg algebra. Thereafter, we have defined Sugawara like construction of the conserved charges and it has been shown that one class of the generators of charges satisfy the Virasoro algebra and the entropy of the surface can be determined from the zeroth mode of charges using the Cardy formula. It reveals the thermodynamic connection of the conserved charges which arise due to the symmetry conditions of the horizon configuration.
 
  As we said, the charges $Q^+$ and $Q^-$ are defined in this work corresponds to $\xi^+$ and $\xi^-$, respectively. Now if we consider only $\xi^+$, it changes our metric  in the following way:  $\delta_{\xi^{+}}g_{ur}=-\p_uF$, $\delta_{\xi^{+}} g_{rr}=0$ and $\delta_{\xi^{+}} g_{uu}=-2r\p_u^2F-r(\p_r\bar{\a})(\p_uF)=\mathcal{O}(r)$; while $\xi^-$ changes as $\delta_{\xi^{-}} g_{ur}=\p_uF$, $\delta_{\xi^{-}} g_{rr}=0$ and $\delta_{\xi^{-}} g_{uu}=2\bar{\a}\p_uF+(\p_u\bar{\a})F=\mathcal{O}(r)$. This shows that the location of the null surface does not change under any one of the diffeomorphism vectors as the null surface is determined by the condition $l^2=g_{ab}l^al^b = g_{uu}=0$ (note, here only $u$ component of $l^a$ is non-zero) and for both cases the leading order correction to $\delta g_{uu}$ is $\mathcal{O}(r)$. It implies the individual vector also respects our original imposed condition to find $\xi^u$ and $\xi^r$.
  
 As mentioned earlier, the study of these algebra on a null surface might enlighten the obscure quantum nature of the gravity as the noncommutativity between the charges (the origin of which is solely determined by the configuration of the null surface) is apparent. It is true that we cannot exactly say whether this quantities, manifesting the noncommutativity, correspond to the physical observables. Nevertheless, our analysis provides the glimpse on the quantum nature that arises due to the configuration of the spacetime. Moreover, the null surface is a local concept and, in general, is not the solution of any equation of motion of the spacetime. Also, it shows all the thermodynamic features. Therefore, in the context of emergent gravity paradigm, where the gravity is considered as the outcome of more fundamental interactions (i.e., thermodynamics), the null surface is highly regarded. Hence, this analysis is important in the theory of emergent gravity as well. 
 
 Let us now comment on the connections and the comparisons with the contemporary works. As mentioned earlier, this type of algebra has also been obtained in the string theory \cite{Seiberg:1999vs}. In that case, the spacetime is itself noncommutative. Moreover, the works which are alike to this one \cite{Afshar:2016wfy, Afshar:2016uax, Grumiller:2016kcp, Sheikh-Jabbari:2016npa, Grumiller:2016pqb, Cai:2016idg, Afshar:2016kjj,Afshar:2017okz, Setare:2016jba, Setare:2016vhy}, shows similar results but, those analysis are confined to the three dimension and the algebra which they show is the usual Heisenberg one. On the contrary, our analysis is valid in any dimensions of the spacetime. Moreover, our algebra casts more light on the quantum nature of the surface as it is noncommutative. 

Some of the open issues are needed to be taken care. In our calculation, the assumption is considered to be as $\alpha$ and $\mu$ are independent of coordinate $u$. Although it appears to be ``OK'' as we have shown that this does not lead to $r=0$ null surface as Killing horizon. In that sense, the present analysis reflects the properties of a generic null surface. But it would be interesting to see if such happens without invoking any assumption as input. Another point has to be noted is that the main results in this paper depends on the specific choice of the Fourier modes of the unknown function $F$, given by (\ref{FF}), seems to be ``add-hoc'' as there is no such, in general, clear periodicity in the coordinates for the metric (\ref{METRIC}). But what is important is that such a choice works well. So one has to look for the justification for this choice. More importantly, we need to see if this is an unique one or there is other choice. It must be mentioned that the same has also been adopted earlier. To tell more on this, in our work, we have taken the periodicity of the $u$ coordinate which is essential for obtaining the compact values of the Fourier charges and brackets. The time coordinate in a Lorentzian manifold, in general, has a periodicity in the Eucledian space. The Schwarzschild metric or even the Rindler metric can be the examples for that. The idea of utilizing the periodicity of the time coordinate in the Fourier mode of the generating function (in our case $F_{m}$) was followed from the earlier work \cite{Silva:2002jq} and was used in several works by one of the authors \cite{Majhi:2011ws, Majhi:2012tf, Majhi:2012nq, Majhi:2013lba, Majhi:2014lka, Majhi:2015tpa}, which was later followed by many other works \cite{Zhang:2012fq, Kim:2013qra, Katsuragawa:2013bma, Chakraborty:2016dwb, Bhattacharya:2017pqc}. Recently, the same idea has been used for the same metric in GNC as well \cite{Chakraborty:2016dwb} while obtaining the Virasoro algebra, where the periodicity of $u$ has been used. So, our assumption of periodicity in the $u$ coordinate is not a bizarre or unfamiliar one. This is widely used in the several works in this line. In our case (or in \cite{Chakraborty:2016dwb}), the periodicity in $u$ is imposed and it is taken to be the same of retarded time coordinate in the Rindler metric. It is because under the certain assumptions in the near null limit, the metric in GNC resembles to the Rindler one. Periodicity of Eucleadian timelike coordinate is common in the discussion of thermodynamics of horizon \cite{Gibbons:1976ue}. Since we are also interested to thermodynamics, it is quite natural to consider such input in this case as well. Such an analysis is semi-classical in nature. Therefore, it can be considered as a mere limitation of the analysis and certainly not as a drawback. Moreover, as mentioned earlier, people find it reasonable to go with this limitation and we too found the same in our work as well. Our intention here is to explore the more deeper meaning of the imposed boundary condition within the existing approaches. But certainly, there is option to improve this.
 
 Our proposal  -- presence of \textit{noncommutative Heisenberg hair} on the horizon -- in this and in the previous work \cite{Majhi:2017fua} implicates that the surface structure the horizon is much ``richer'' than what was predicted by the earlier works. The whole result is general and can be fitted in any theory of gravity. Hope we can provide more insights in this regards in near future.
 
\vskip 4mm
{\section*{Acknowledgments}}
\noindent 
We thank T. Padmanabhan for critical and useful comments during the execution of the work. Also we thank Sumanta Chakraborty for clarifying some of the issues. The anonymous referee is greatly acknowledged for giving several constructive and important comments which help us to greatly improve the earlier version. The research of one of the authors (BRM) is supported by a START-UP RESEARCH GRANT (No. SG/PHY/P/BRM/01) from Indian Institute of Technology
Guwahati, India.
\vskip 5mm
\appendix
\section{Obtaining $\xi^a$ in Eq. \eqref{XIUP} } \label{APPNEW}
Let us consider the first condition $\lie g_{rr}=0$, which implies $\na_r\xi_r=0$. Hence 
\begin{align}
g_{ri}\na_r\xi^i=0~. \label{mid1}
\end{align}
One can write \eqref{mid1} further as
\begin{align}
\p_r\xi^u+\G^u_{rr}\xi^r+\G^u_{ur}\xi^u=0~.
\end{align}
Since $\G^u_{rr}=\G^u_{ur}=0$ (see \cite{MORALES}), one can obtain the expression of $\xi^u$ given in the Eq. \eqref{XIUP}.

We now take the second condition, $\lie g_{ru}=0$.  It implies $g_{ui}\na_r\xi^i+g_{ri}\na_u\xi^i=0$, which further results in 
\begin{align}
g_{uu}[\p_r\xi^u+\G^u_{ri}\xi^i]+g_{ur}[\p_r\xi^r+\G^r_{ri}\xi^i]+g_{ru}[\p_u\xi^u
\no 
\\
+\G^u_{ui}\xi^i]+g_{uA}[\p_r\xi^A+\G^A_{ri}\xi^i]=0~. \label{mid2}
\end{align}
If one substitutes the values of $\xi^u$ and of all $\G$'s from \cite{MORALES}, the final expression of \eqref{mid2} reduces to $\p_r\xi^r+\p_uF=0$, from which one can obtain the expression of $\xi^r$ given in Eq. \eqref{XIUP}.
%
\section{Killing horizon?} \label{APP2}
In our analysis, we have taken $\a$ and $\mu$ being independent of the coordinate $u$ only on the null surface, as the leading order contribution of these quantities are considered to be independent of $u$. Besides, the null surface $r=0$ in our case is defined by the vanishing of the norm of the vector $l^a=(\p/\p u)^a$. So the natural question comes in one's mind: Are the imposed conditions lead to the fact that the null surface, in our analysis, is a Killing horizon? However, we shall show that $l^a$ does not define a Killing horizon.

For our present metric (\ref{METRIC}), one can check that 
\begin{align}
\pounds_lg_{ab}=\p_ug_{ab}~.
\end{align}
At the first glance, one might be tempted to remark that our assumption, that the leading order contribution of $\a$ and $\mu$ are independent of $u$, might lead to the fact that $l^a$ is a Killing vector on the null surface. However, this is not the case. It must be noted that not all the metric components are needed to be independent of $u$ and therefore right hand side of the above does not vanish for all values of indices $a$ and $b$. Let us explain the reason little elaborately. As it has been mentioned earlier, we have taken the determinant of transverse metric $\mu$ being independent of $u$; not all the components of $\mu_{AB}$. In this situation one can analytically verify that $\mu$ can be independent of $u$ even though $\mu_{AB}$ depends on it.
Secondly, $l^a$ does not satisfy the Frobenius theorem as it is not hypersurface orthogonal all over the spacetime. 
For this reasons, we can say that our assumptions do not imply the null surface, defined by $l^2=0$, to be a Killing one. 

One might be interested to seek if there is any other possible Killing vector, satisfying the Frobenius theorem, whose vanishing norm defines our $r=0$ null horizon as a Killing one. The answer will be ``No''. We have taken $\a$ and $\mu$ to be independent of $u$ only and not of any other coordinate. Therefore, we cannot imagine that the Killing vector to be a linear sum of two vectors (as is the case of a Kerr black hole, where the metric admits two Killing vector, and the horizon surface is determined by the linear combination of the two Killing vectors). We have checked that $l^a=(\p/\p u)^a$ is not a Killing vector and it does not define any Killing horizon. Therefore, any vector proportional to $l^a$ will not define a Killing horizon as well.

 To sum up, the null surface is not assumed to be a Killing one when the assumption -- $\a$ and $\mu$ are independent of $u$ -- is taken. In that sense the whole analysis predicts the properties of a generic null surface.
 \vskip 5mm
\section{To prove (\ref{QFOU}) and (\ref{BRACFOU})} \label{APP}
The Fourier modes of the integrations, (\ref{Q+}), (\ref{Q-}), (\ref{RES1}), (\ref{RES2}) and (\ref{RES3}), which are needed to be evaluated to obtain the Fourier modes of the charges and of the brackets in the desired form, are highly daunting task as the functional form of $\a$ and $\mu$ are not exactly known. In an earlier attempt \cite{Chakraborty:2016dwb}, as we have mentioned in our main analysis, assumed that $\a$ and the transverse metric coefficients $\mu_{AB}$ are independent of $u$ and a particular transverse coordinate. Here, we find that a much more weaker restriction can be imposed. It will be shown that taking the leading order terms of $\a$ and $\mu$ (not the transverse metric coefficients) as independent of \textit{only $u$ coordinate is enough} to get the near horizon algebra. In our case, the periodicity condition of $u$ helps to get the desired results even if the functional form of $\a$ and $\mu$ are not known. We have also found out the physical interpretation of $\mu$ being independent of $u$ which is mentioned earlier. In the next appendix, it will be shown that this assumption does not imply a Killing horizon near to the null surface.
 
 Now, while calculating the Fourier modes from the Eqns. \eqref{Q+}, \eqref{Q-}, \eqref{RES1}, \eqref{RES2} and \eqref{RES3} with the Fourier mode of $F$ as mentioned earlier, we mainly encounter two integrations, i.e., $I_1\sim\int\sqrt{\mu}d^{d-2}x^Ae^{i(m+n)(au+p_Ax^A)}$ and $I_2\sim\int\sqrt{\mu}d^{d-2}x^A\a e^{i(m+n)(au+p_Ax^A)}$. To solve these two integrations, we have used the periodicity condition on the coordinate $u$ with the fact that the periodicity is independent of any coordinate. To get the value of the integration $I_1$, let us start with the fact that
\begin{align}
 \int_0^Rdu\int d^{d-2}x^A\sqrt{\mu}e^{i(m+n)(au+p_Ax^A)}
 \no 
 \\
 =R\int\sqrt{\mu}d^{d-2}x^A \d_{m+n,0}~.\ \ \ \ \ \ \ \label{UINT}
\end{align}
In the above we have used the result $\int_0^Rdue^{i(m+n)au}=R\delta_{m+n,0}$ and $R=2\pi/a$.
The next step is to take derivative with respect to $R$ on the both sides of (\ref{UINT}). From the left hand side we get
\begin{align}
\int\sqrt{\mu}d^{d-2}x^A\frac{\p}{\p R}\int_0^Rdue^{i(m+n)(au+p_Ax^A)} \ \
\no 
\\
=\int\sqrt{\mu}d^{d-2}x^Ae^{i(m+n)(aR+p_Ax^A)} \ \ \ \
\no 
\\
=\int\sqrt{\mu}d^{d-2}x^Ae^{i(m+n)p_Ax^A}~, \ \ \ \ \ \ \label{A2}
\end{align}
where, in the second equality $\frac{\p}{\p b}[\int_a^bf(x)dx]=f(b)$ has been used and the final result is obtained after substitution of $R=2\pi/a$. Whereas, the derivative with respect to $R$ on the right hand side of (\ref{UINT}) gives $\int\sqrt{\mu}d^{d-2}x^A \d_{m+n,0}$. This implies
\begin{align}
\int\sqrt{\mu}d^{d-2}x^Ae^{i(m+n)(au+p_Ax^A)}=A\d_{m+n, 0} \label{MAIN1}
\end{align}
 where $A=\int\sqrt{\mu}d^{d-2}x^A$ is the transverse surface area.

The second integration $I_2$ can be performed following the previous method of $I_1$ and considering $\a$ as independent of $u$ while performing the integration of $u$.  One can similarly obtain
\begin{align}
\int\sqrt{\mu}d^{d-2}x^A\a e^{i(m+n)(au+p_Ax^A)}=\a A\d_{m+n, 0}~. \label{MAIN2}
\end{align}
Using the results of \eqref{MAIN1} and \eqref{MAIN2}, the Fourier modes of the charges and the brackets of the charges can be obtained which are given in \eqref{QFOU} and \eqref{BRACFOU}. 



\begin{thebibliography}{99}
\bibitem{Bekenstein:1973ur} 
  J.~D.~Bekenstein,
  ``Black holes and entropy,''
  Phys.\ Rev.\ D {\bf 7}, 2333 (1973).
  
\bibitem{Hawking:1974sw} 
  S.~W.~Hawking,
  ``Particle Creation by Black Holes,''
  Commun.\ Math.\ Phys.\  {\bf 43}, 199 (1975)
  Erratum: [Commun.\ Math.\ Phys.\  {\bf 46}, 206 (1976)].
  
\bibitem{Bardeen:1973gs} 
  J.~M.~Bardeen, B.~Carter and S.~W.~Hawking,
  ``The Four laws of black hole mechanics,''
  Commun.\ Math.\ Phys.\  {\bf 31}, 161 (1973).
  
\bibitem{Parattu:2013gwa} 
  K.~Parattu, B.~R.~Majhi and T.~Padmanabhan,
  ``Structure of the gravitational action and its relation with horizon thermodynamics and emergent gravity paradigm,''
  Phys.\ Rev.\ D {\bf 87}, no. 12, 124011 (2013)
  [arXiv:1303.1535 [gr-qc]].
 
\bibitem{Chakraborty:2015aja} 
  S.~Chakraborty, K.~Parattu and T.~Padmanabhan,
  ``Gravitational field equations near an arbitrary null surface expressed as a thermodynamic identity,''
  JHEP {\bf 1510}, 097 (2015)
  [arXiv:1505.05297 [gr-qc]].
  
\bibitem{Chakraborty:2015hna} 
  S.~Chakraborty and T.~Padmanabhan,
  ``Thermodynamical interpretation of the geometrical variables associated with null surfaces,''
  Phys.\ Rev.\ D {\bf 92}, no. 10, 104011 (2015)
  [arXiv:1508.04060 [gr-qc]].
  
\bibitem{Padmanabhan:2009vy} 
  T.~Padmanabhan,
  ``Thermodynamical Aspects of Gravity: New insights,''
  Rept.\ Prog.\ Phys.\  {\bf 73}, 046901 (2010)
  [arXiv:0911.5004 [gr-qc]].
  
\bibitem{Majhi:2011ws} 
  B.~R.~Majhi and T.~Padmanabhan,
  ``Noether Current, Horizon Virasoro Algebra and Entropy,''
  Phys.\ Rev.\ D {\bf 85}, 084040 (2012)
  [arXiv:1111.1809 [gr-qc]].
  
\bibitem{Majhi:2012tf} 
  B.~R.~Majhi and T.~Padmanabhan,
  ``Noether current from the surface term of gravitational action, Virasoro algebra and horizon entropy,''
  Phys.\ Rev.\ D {\bf 86}, 101501 (2012)
  [arXiv:1204.1422 [gr-qc]].
  
\bibitem{Majhi:2012nq} 
  B.~R.~Majhi,
  ``Noether current of the surface term of Einstein-Hilbert action, Virasoro algebra and entropy,''
  Adv.\ High Energy Phys.\  {\bf 2013}, 386342 (2013)
  [arXiv:1210.6736 [gr-qc]].
 
 \bibitem{Majhi:2013lba} 
  B.~R.~Majhi and S.~Chakraborty
  ``Anomalous effective action, Noether current, Virasoro algebra and Horizon entropy,''
  Eur.\ Phys.\ J.\ C {\bf 74}, 2867 (2014)
  [arXiv:1311.1324 [gr-qc]].
  
 \bibitem{Majhi:2014lka} 
  B.~R.~Majhi,
  ``Conformal Transformation, Near Horizon Symmetry, Virasoro Algebra and Entropy,''
  Phys.\ Rev.\ D {\bf 90}, no. 4, 044020 (2014)
  [arXiv:1404.6930 [gr-qc]].
  
\bibitem{Majhi:2015tpa} 
  B.~R.~Majhi,
  ``Near horizon hidden symmetry and entropy of Sultana-Dyer black hole: A time dependent case,''
  Phys.\ Rev.\ D {\bf 92}, no. 6, 064026 (2015)
  [arXiv:1505.03310 [gr-qc]].
  
\bibitem{Bhattacharya:2017pqc} 
  K.~Bhattacharya and B.~R.~Majhi,
  ``Fresh look at the scalar-tensor theory of gravity in Jordan and Einstein frames from undiscussed standpoints,''
  Phys.\ Rev.\ D {\bf 95}, no. 6, 064026 (2017)
  [arXiv:1702.07166 [gr-qc]].
  
\bibitem{Chakraborty:2016dwb} 
  S.~Chakraborty, S.~Bhattacharya and T.~Padmanabhan,
  ``Entropy of a generic null surface from its associated Virasoro algebra,''
  Phys.\ Lett.\ B {\bf 763}, 347 (2016)
  [arXiv:1605.06988 [gr-qc]].
  
\bibitem{francesco}
 P.~Di~Francesco, P.~Mathieu, D.~S\'en\'echal, \textit{Conformal Field Theory} (Springer-Verlag, New York, 1997).
 
\bibitem{Cardy:1986ie} 
  J.~L.~Cardy,
  ``Operator Content of Two-Dimensional Conformally Invariant Theories,''
  Nucl.\ Phys.\ B {\bf 270}, 186 (1986).

\bibitem{Brown:1986nw} 
  J.~D.~Brown and M.~Henneaux,
  ``Central Charges in the Canonical Realization of Asymptotic Symmetries: An Example from Three-Dimensional Gravity,''
  Commun.\ Math.\ Phys.\  {\bf 104}, 207 (1986).
  
\bibitem{Carlip:1998wz} 
  S.~Carlip,
  ``Black hole entropy from conformal field theory in any dimension,''
  Phys.\ Rev.\ Lett.\  {\bf 82}, 2828 (1999)
  [hep-th/9812013].

 \bibitem{Carlip:1999cy} 
  S.~Carlip,
  ``Entropy from conformal field theory at Killing horizons,''
  Class.\ Quant.\ Grav.\  {\bf 16}, 3327 (1999)
  [gr-qc/9906126].  
  
\bibitem{Majhi:2017fua} 
  B.~R.~Majhi,
  ``Noncommutativity in near horizon symmetries in gravity,''
  Phys.\ Rev.\ D {\bf 95}, no. 4, 044020 (2017)
  [arXiv:1701.07952 [gr-qc]].
  
\bibitem{Jacobson:1995ab} 
  T.~Jacobson,
  ``Thermodynamics of space-time: The Einstein equation of state,''
  Phys.\ Rev.\ Lett.\  {\bf 75}, 1260 (1995)
  [gr-qc/9504004].
  
  \bibitem{Majhi:2012st} 
  B.~R.~Majhi,
  ``Gravitational anomalies and entropy,''
  Gen.\ Rel.\ Grav.\  {\bf 45}, 345 (2013)
  [arXiv:1210.3306 [gr-qc]].
  
\bibitem{Seiberg:1999vs} 
  N.~Seiberg and E.~Witten,
  ``String theory and noncommutative geometry,''
  JHEP {\bf 9909}, 032 (1999)
  [hep-th/9908142].
  
\bibitem{Bondi:1962px} 
  H.~Bondi, M.~G.~J.~van der Burg and A.~W.~K.~Metzner,
  ``Gravitational waves in general relativity. 7. Waves from axisymmetric isolated systems,''
  Proc.\ Roy.\ Soc.\ Lond.\ A {\bf 269}, 21 (1962).
  
\bibitem{Sachs:1962wk} 
  R.~K.~Sachs,
  ``Gravitational waves in general relativity. 8. Waves in asymptotically flat space-times,''
  Proc.\ Roy.\ Soc.\ Lond.\ A {\bf 270}, 103 (1962).
  
\bibitem{Sachs:1962zza} 
  R.~Sachs,
  ``Asymptotic symmetries in gravitational theory,''
  Phys.\ Rev.\  {\bf 128}, 2851 (1962).
  
\bibitem{Afshar:2016wfy} 
  H.~Afshar, S.~Detournay, D.~Grumiller, W.~Merbis, A.~Perez, D.~Tempo and R.~Troncoso,
  ``Soft Heisenberg hair on black holes in three dimensions,''
  Phys.\ Rev.\ D {\bf 93}, no. 10, 101503 (2016)
  [arXiv:1603.04824 [hep-th]].
  
\bibitem{Afshar:2016uax} 
  H.~Afshar, D.~Grumiller and M.~M.~Sheikh-Jabbari,
  ``Black Hole Horizon Fluffs: Near Horizon Soft Hairs as Microstates of Three Dimensional Black Holes,''
  arXiv:1607.00009 [hep-th].
  
\bibitem{Grumiller:2016kcp} 
  D.~Grumiller, A.~Perez, S.~Prohazka, D.~Tempo and R.~Troncoso,
  ``Higher Spin Black Holes with Soft Hair,''
  JHEP {\bf 1610}, 119 (2016)
  [arXiv:1607.05360 [hep-th]].
  
\bibitem{Sheikh-Jabbari:2016npa} 
  M.~M.~Sheikh-Jabbari and H.~Yavartanoo,
  ``Horizon Fluffs: Near Horizon Soft Hairs as Microstates of Generic $AdS_3$ Black Holes,''
  Phys.\ Rev.\ D {\bf 95}, no. 4, 044007 (2017)
  [arXiv:1608.01293 [hep-th]].
  
\bibitem{Grumiller:2016pqb} 
  D.~Grumiller and M.~Riegler,
  ``Most general AdS$_{3}$ boundary conditions,''
  JHEP {\bf 1610}, 023 (2016)
  [arXiv:1608.01308 [hep-th]].
  
\bibitem{Cai:2016idg} 
  R.~G.~Cai, S.~M.~Ruan and Y.~L.~Zhang,
  ``Horizon supertranslation and degenerate black hole solutions,''
  JHEP {\bf 1609}, 163 (2016)
  [arXiv:1609.01056 [gr-qc]].
  
\bibitem{Afshar:2016kjj} 
  H.~Afshar, D.~Grumiller, W.~Merbis, A.~Perez, D.~Tempo and R.~Troncoso,
  ``Soft hairy horizons in three spacetime dimensions,''
  Phys.\ Rev.\ D {\bf 95}, no. 10, 106005 (2017)
  [arXiv:1611.09783 [hep-th]].
  
\bibitem{Afshar:2017okz} 
  H.~Afshar, D.~Grumiller, M.~M.~Sheikh-Jabbari and H.~Yavartanoo,
  ``Horizon fluff, semi-classical black hole microstates - Log-corrections to BTZ entropy and black hole/particle correspondence,''
  arXiv:1705.06257 [hep-th].
  
\bibitem{Setare:2016jba} 
  M.~R.~Setare and H.~Adami,
  ``Near Horizon Symmetries of the Non-Extremal Black Hole Solutions of Generalized Minimal Massive Gravity,''
  Phys.\ Lett.\ B {\bf 760}, 411 (2016)
  [arXiv:1606.02273 [hep-th]].
  
\bibitem{Setare:2016vhy} 
  M.~R.~Setare and H.~Adami,
  ``The Heisenberg algebra as near horizon symmetry of the black flower solutions of Chern -- Simons -- like theories of gravity,''
  Nucl.\ Phys.\ B {\bf 914}, 220 (2017)
  [arXiv:1606.05260 [hep-th]].
 
\bibitem{Hollands:2006rj} 
  S.~Hollands, A.~Ishibashi and R.~M.~Wald,
  ``A Higher dimensional stationary rotating black hole must be axisymmetric,''
  Commun.\ Math.\ Phys.\  {\bf 271}, 699 (2007)
  [gr-qc/0605106].
  
  \bibitem{MORALES}
  E.~M.~Morales,
  ``On a Second Law of Black Hole Mechanics in a Higher Derivative Theory of Gravity,''
  http://www.theorie.physik.uni-goettingen.de/forschung/qft/theses/dipl/Morfa-Morales.pdf
  
 \bibitem{Moncrief:1983xua} 
  V.~Moncrief and J.~Isenberg,
  ``Symmetries of cosmological Cauchy horizons,''
  Commun.\ Math.\ Phys.\  {\bf 89}, no. 3, 387 (1983).
  
\bibitem{Eling:2012xa} 
  C.~Eling, A.~Meyer and Y.~Oz,
  ``Local Entropy Current in Higher Curvature Gravity and Rindler Hydrodynamics,''
  JHEP {\bf 1208}, 088 (2012)
  [arXiv:1205.4249 [hep-th]].
  
\bibitem{Eling:2016xlx} 
  C.~Eling and Y.~Oz,
  ``On the Membrane Paradigm and Spontaneous Breaking of Horizon BMS Symmetries,''
  JHEP {\bf 1607}, 065 (2016)
  [arXiv:1605.00183 [hep-th]].
  
\bibitem{Iyer:1994ys} 
  V.~Iyer and R.~M.~Wald,
  ``Some properties of Noether charge and a proposal for dynamical black hole entropy,''
  Phys.\ Rev.\ D {\bf 50}, 846 (1994)
  [gr-qc/9403028].
  
\bibitem{Padmanabhan:2013xyr} 
  T.~Padmanabhan and D.~Kothawala,
  ``Lanczos-Lovelock models of gravity,''
  Phys.\ Rept.\  {\bf 531}, 115 (2013)
  [arXiv:1302.2151 [gr-qc]].
  
\bibitem{Donnay:2015abr} 
  L.~Donnay, G.~Giribet, H.~A.~Gonzalez and M.~Pino,
  Phys.\ Rev.\ Lett.\  {\bf 116}, no. 9, 091101 (2016)
  doi:10.1103/PhysRevLett.116.091101
  [arXiv:1511.08687 [hep-th]].
  
\bibitem{Silva:2002jq} 
  S.~Silva,
  ``Black hole entropy and thermodynamics from symmetries,''
  Class.\ Quant.\ Grav.\  {\bf 19}, 3947 (2002)
  [hep-th/0204179].
  
    
\bibitem{Donnay:2016ejv} 
  L.~Donnay, G.~Giribet, H.~A.~González and M.~Pino,
  ``Extended Symmetries at the Black Hole Horizon,''
  JHEP {\bf 1609}, 100 (2016)
  [arXiv:1607.05703 [hep-th]].
  
  
\bibitem{Nielsen:2008cr} 
  A.~B.~Nielsen,
  ``Black holes and black hole thermodynamics without event horizons,''
  Gen.\ Rel.\ Grav.\  {\bf 41}, 1539 (2009)
  [arXiv:0809.3850 [hep-th]].
  
  
\bibitem{Booth:2012xm} 
  I.~Booth,
  ``Spacetime near isolated and dynamical trapping horizons,''
  Phys.\ Rev.\ D {\bf 87}, no. 2, 024008 (2013)
  [arXiv:1207.6955 [gr-qc]].
  
\bibitem{Zhang:2012fq} 
  S.~J.~Zhang and B.~Wang,
  ``Surface term, Virasoro algebra and Wald entropy of black holes in higher curvature gravity,''
  Phys.\ Rev.\ D {\bf 87}, no. 4, 044041 (2013)
  [arXiv:1212.6896 [hep-th]].
  
\bibitem{Kim:2013qra} 
  W.~Kim, S.~Kulkarni and S.~H.~Yi,
  ``Conserved quantities and Virasoro algebra in New massive gravity,''
  JHEP {\bf 1305}, 041 (2013)
  [arXiv:1303.3691 [hep-th]].
  
\bibitem{Katsuragawa:2013bma} 
  T.~Katsuragawa and S.~Nojiri,
  ``Noether current from surface term, Virasoro algebra and black hole entropy in bigravity,''
  Phys.\ Rev.\ D {\bf 87}, no. 10, 104032 (2013)
  [arXiv:1304.3181 [hep-th]].
  
\bibitem{Gibbons:1976ue} 
  G.~W.~Gibbons and S.~W.~Hawking,
  ``Action Integrals and Partition Functions in Quantum Gravity,''
  Phys.\ Rev.\ D {\bf 15}, 2752 (1977).
  
  \end{thebibliography}
\end{document}